\documentstyle[preprint,aps]{revtex}
\newcommand{\be}{\begin{eqnarray}}
\newcommand{\ee}{\end{eqnarray}}
\newcommand{\ra}{\rightarrow}
\begin{document}
\preprint{
YUMS 95--11
\hspace{-42mm}
\raisebox{2.5ex}{
KAIST-CHEP-95/04
}
\hspace{-32.5mm}
\raisebox{-2.5ex}{SNUTP 95--039}
}
\title{ Form Factors for Exclusive Semileptonic $B$--Decays }
\vspace{0.65in}
\author{
C. S. ~Kim\thanks{kim@cskim.yonsei.ac.kr} }
\vspace{.5in}
\address{
Department of Physics, Yonsei University, Seoul 120--749, KOREA}
\vspace{0.25in}
\author{ Jae Kwan ~Kim,~
Yeong Gyun ~Kim\thanks{ygkim@chep6.kaist.ac.kr}~ and~
Kang Young ~Lee\thanks{kylee@chep5.kaist.ac.kr}
}
\vspace{.5in}
\address{
Department of Physics, KAIST, Taejon 305--701, KOREA}
\date{\today}
\maketitle
\begin{abstract}
\\
We developed the new parton model approach for
exclusive semileptonic decays of $B$-meson to $D,~D^*$ by
extending the inclusive parton model,
and by combining with the results of the HQET, motivated by
Drell-Yan process.
Without the nearest pole dominance ans\"atze, we {\bf derived} the
dependences of hadronic form factors on $q^2$. We also calculated
numerically the slope of the Isgur-Wise function, which is consistent
with the experimental results.
\end{abstract}
\pacs{ }

\narrowtext

\section{Introduction}
In exclusive weak decay processes of hadrons, the effects of strong
interaction are encoded in hadronic form factors.
These decay form factors are Lorentz invariant
functions which depend on the momentum transfer $q^2$,
and their behaviors with varying $q^2$ are dominated by non-perturbative
effects of QCD.

$B$--meson decay processes have been studied in detail
as providing many interesting informations on the interplay
of electroweak and strong interactions and as a source extracting
the parameters of weak interactions, such as $|V_{cb}|$ and $|V_{ub}|$.
As more data will be accumulated from the asymmetric $B$--factories
in near future,
the theoretical and experimental studies on exclusive $B$--meson decays
would also give better understandings on the Standard Model and its
possible extensions.

Over the past few years, a great progress has been achieved
in our understanding of the exclusive semileptonic decays
of heavy flavors to heavy flavors \cite{neubert}.
In the limit where the mass of the heavy quark is taken to infinity,
its stong interactions become independent of its mass and spin,
and depend only on its velocity.
This provides a new $SU(2N_f)$ spin--flavor symmetry,
which is not manifest in the theory of QCD. However,
this new symmetry has been made explicit
in a framework of the heavy quark effective theory (HQET) \cite{hqet}.
In practice, the HQET and this new symmetry relate
all the hadronic matrix elements of
$B \ra D$ and $B \ra D^*$ semileptonic decays,
and all the form factors can be reduced to a single universal function,
the so-called Isgur-Wise function \cite{hqet,isgurfun},
which represents the common non-perturbative dynamics of
weak decays of heavy mesons.
However, the HQET cannot predict the values of the Isgur-Wise function
over the whole $q^2$ range, though the normalization of the Isgur-Wise
function is precisely known in the zero recoil limit.
Hence the extrapolation of $q^2$ dependences of the Isgur-Wise function
and of all form factors is still model dependent and the source of
uncertainties in any theoretical model.
Therefore, it is strongly recommended to determine
hadronic form factors of $B$--meson decay more reliably, when we think
of their importance in theoretical and experimental analyses.

In this paper we developed the parton model approach {\bf for exclusive
semileptonic} $B$ decays to $D,~D^*$, and predicted the $q^2$ dependences
of all form factors. Previously the parton model approach has been
established to describe inclusive semileptonic $B$ decays \cite{pas,jin},
and found to give excellent agreements with experiments for electron
energy spectrum at all energies.
While many attempts describing exclusive $B$ decays often take
the pole-dominance ans$\ddot{\mbox {a}}$tze
as behaviors of form factors with varying $q^2$ \cite{wsb,ks},
in our approach they are derived by the kinematical relations
between initial $b$ quark and final $c$ quark.
According to the Wirbel {\it et. al.} model \cite{wsb},
which is one of the most popular
model to describe exclusive decays of $B$ mesons,
the hadronic form factors are related to
the meson wavefunctions' overlap-integral in the infinite momentum
frame, but in our model they are determined by integral
of the fragmentation functions, which are experimentally measuable.


For completeness, here we briefly review the parton model approach
for inclusive semileptonic decays of $B$ meson.
The parton model approach pictures the mesonic decay
as the decay of the partons
in analogy to deep inelastic scattering process.
The probability of finding a $b$-quark in a $B$ meson carrying a fraction
$x$ of the meson momentum in the infinite momentum frame
is given by the distribution function $f(x)$.
Then we write the Lorentz invariant decay width as follows:
\be
E_B~d\Gamma(B \ra X_q e \nu) = \int~dx~f(x)~
                 E_b~d\Gamma(b \ra qe\nu)~~,
\ee
with the relation $p_b = x p_B$.
Using standard definitions of the structure functions
for the hadronic tensor $W^{\mu \nu}$
\be
W^{\mu \nu} &=& -g^{\mu \nu}~W_1 + p_B^{\mu} p_B^{\nu}~\frac{W_2}{m_B^2}
   -i \epsilon^{\mu \nu \alpha \beta} {p_B}_{\alpha} q_{\beta}~
  \frac{W_3}{2 m_B^2} \nonumber \\
            && + q^{\mu} q^{\nu}~\frac{W_4}{m_B^2}
	+(p_B^{\mu} q^{\nu} + p_B^{\nu} q^{\mu})~\frac{W_5}{2 m_B^2}
	+i(p_B^{\mu} q^{\nu} - p_B^{\nu} q^{\mu})~\frac{W_6}{2 m_B^2}~~,
\ee
we obtain
\be
&&W_1 = \frac{1}{2} \left(f(x_{+})+f(x_{-})\right)~~, \nonumber \\
&&\frac{W_2}{m_B^2} = \frac{2}{m_B^2 (x_+-x_-)}
		     \left( x_{+}f(x_{+}) - x_{-}f(x_{-}) \right)~~,
            \nonumber \\
&&\frac{W_3}{2 m_B^2} = -\frac{1}{m_B^2 (x_+-x_-)}
		       \left( f(x_{+}) - f(x_{-}) \right)~~,
            \nonumber \\
&& \mbox{and}~~~~~~W_4 = W_5 = W_6 = 0~~,  \nonumber
\ee
where
\be
x_{\pm} = \frac{q_0 \pm \sqrt{|{\bf q}|^2+m_q^2}}{m_B}~~, \nonumber
\ee
with the final state quark mass $m_q$.
Hence the double differential decay rate is given by
\be
\frac{d\Gamma}
     {dE_e dy dx_+} = \frac{G^2_F M^4_B |V_{qb}|^2}
                           {8 \pi^3}
y \{x_+ f(x_+) (x_+-y-m_q^2/m_B^2/x_+) -(x_+ \leftrightarrow x_-) \}~~,
\ee
where $y=2E_{\nu}/m_B$. In the limit $f(z)= \delta(1-z)$, we
reproduce the HQET leading term except that $m_b$ is replaced
by $m_B$.
In Section II, we develop the parton model approach
for exclusive semileptonic
decays of $B$ meson, and give all the theoretical details for
$B \ra D l \nu$ and $B \ra D^* l \nu$. Section III contains discussions
and conclusions of this paper.

\section{Parton Model Approach for Exclusive Decays of B Meson}

We now develop the parton model approach for
exclusive semileptonic decays of $B$ meson by
extending the previously explained inclusive parton model,
and by combining with the results of the HQET.
Theoretical formulation of this approach is, in a sence, closely related to
Drell-Yan process, while the parton model of inclusive $B$ decays is
motivated by deep inelastic scattering process.
And the bound state effects of exclusive $B$ decays are encoded into
the hadronic distribution functions of partons inside an initial $B$ meson
and  of partons of a final state resonance hadron.
Then, the Lorentz invariant hadronic decay width can be obtained
using the structure functions, as in Eq. (1),
\be
E_B \cdot d\Gamma(B \ra D(D^*)e\nu) = \int dx \int dy~
                      f_B(x)~E_b \cdot d\Gamma(b \ra ce\nu)~f_D(y)~~.
\ee
The first integral represents the effects of motion of $b$ quark
within $B$ meson and the second integral those of $c$ quark within $D$ meson.
The variables $x$ and $y$ are fractions of momenta of partons
to momenta of mesons,
\be
p_b = x p_B~,~~~~~~~~~
p_c = y p_D~,
\ee
in the infinite momentum frame.
The functions $f_B(x)$ and  $f_D(y)$ are the distribution function
of $b$ quark inside $B$ meson, and the fragmentation function of $c$ quark
to $D$ meson respectively.
Since the momentum fractions and the distribution functions are all defined
in the infinite momentum frame, we have to consider the
Lorentz invariant quantity, ~$E \cdot d\Gamma$ as defined in Eq. (1)
and (4), to use at any other frame.

For a heavy quark ($Q=t,b,c$) the distribution and fragmentation functions
in a heavy meson ($Q q$),
which are closely related by a time reversal transformation,
are of similar functional forms, and peak both at large value of $x$.
Therefore, we follow the previous work of Paschos {\it et. al.} \cite{pas}
to use the Peterson's fragmentation function  \cite{peterson} for both
distributions,  $f_B(x)$ and  $f_D(y)$.
It has the functional form:
\be
f_Q(z) = N z^{-1}~
     \left( 1-\frac{1}{z}-\frac{\epsilon_Q}{1-z} \right)^{-2}~~,
\ee
where $N$ is a normalization constant, and $Q$ denotes $b$ or $c$ quark.

In the Drell-Yan process, the rest degrees of freedom of initial
nucleons which do not take part in the scattering
make incoherent final states, see in Fig. 1 (a).
In the exclusive semileptonic decay of a heavy meson into a final state
heavy meson, however, two sets of left-over light-degrees of freedom are
summed to have the connection,
\be
\left| <D({\rm or}~D^*)|J|B> \right|^2 \sim
                  \sum_{\mbox{{\tiny spin}}, X_1, X_2}
                  \big|~\{<X_2|<c|\}~J~\{|b>|X_1>\}~\big|^2~~.
\ee
For more explicit meaning of Eq. (7), see Fig. 1 (b) and
Section II A and B.
Two sets of states, $ |X_1>$, $|X_2>$ are not independent here,
in fact. And to connect them we need a relation between $x$ and $y$
from the decay kinematics.
The momentum transfer of the decay between mesons is defined as
\be
q \equiv p_B - p_D~~.
\ee
On the other hand, the momentum transfer of the partonic subprocess
is given by
\be
q^{\mbox{{\tiny (parton)}}} = p_b - p_c = x p_B - y p_D~~.
\ee
In fact, the heavy meson's momentum would be $p_H=p_Q+k+{\cal O}
(1/m_Q)$, where $H=B,D$ or $D^*$, and  $Q=b$ or $c$.
And $k$ denotes the momentum of the light-degrees of freedom,
and is related to the effective  mass of a common light quark,
$\bar \Lambda$.
Therefore we have $q = q^{\mbox{{\tiny (parton)}}}$ up to
the common part of the $1/m_Q$ corrections.
With these kinematic relations we derived the following relation
\be
y(x,q^2) = \frac{1}{m_D} \sqrt{ x(x-1) m_B^2 + (1-x) q^2 + x m_D^2}~~.
\ee
Substituting $y$ of Eq. (4) for $y(x)$ of Eq. (10), the double integral
of Eq. (4) is reduced to the single integral over $x$.
Using this relation, we can sum the intermediate states
in Fig. 1(b), as in (7). In Fig. 1, we show the schematic diagrams of
Drell-Yan process and the related exlusive semileptonic decay of $B$ meson.

We note here that the connection (7) and the kinematic relation (10) are
valid approximations for the heavy-to-heavy resonance decays,
with the common light-degrees of freedom of the size ${\cal O}(1/m_Q)$.
As explained before, in the limit where $f_Q(x)= \delta(1-x)$ by increasing
$m_Q$ to infinity, we can reproduce the HQET leading term.
By comparison, the inclusive parton model approach is more reliable  for
the heavy-to-light non-resonant decays to final states of many particles.

\subsection{$B \ra D e \nu$}

{}From Lorentz invariance we write the matrix element of the decay
$\bar{B} \ra D e \bar{\nu}$ in the form
\be
<D|J_{\mu}|B> = f_+(q^2) (p_B+p_D)_{\mu} + f_-(q^2) (p_B-p_D)_{\mu}~~,
\ee
and in terms of the HQET
\be
<D(v')|J_{\mu}|B(v)> = \sqrt{m_B m_D}~(\xi_+(v \cdot v') (v+v')_{\mu}
				     + \xi_-(v \cdot v') (v-v')_{\mu})~~.
\ee
Due to the conservation of leptonic currents, the form factors
multiplicated by $q_{\mu}$ do not contribute.
Then the hadronic tensor is given by
\be
H_{\mu \nu} &=& <D|J_{\mu}|B><D|J_{\nu}|B>^*
           \nonumber \\
            &=& 2~|f_+(q^2)|^2 ({p_B}_{\mu} {p_D}_{\nu}
                              + {p_B}_{\nu} {p_D}_{\mu})~~,
\ee
and can be expressed by the Isgur-Wise function,
\be
H_{\mu \nu} = R^{-1} |\xi(v \cdot v')|^2
             ({p_B}_{\mu} {p_D}_{\nu} + {p_B}_{\nu} {p_D}_{\mu})
                       \left( 1+{\cal O}(\frac{1}{m_Q}) \right)~~,
\ee
where
\be
R=\frac{2\sqrt{m_B m_D}}{m_B+m_D}~~. \nonumber
\ee

The partonic subprocess decay width is given by
\be
E_b \cdot d\Gamma (b \ra ce\nu) &=&
             \frac{1}{2} (2\pi)^4 \delta^4(p_b-p_c-q) \cdot
            2 {G_F}^2 |V_{cb}|^2 H_{\mu \nu}^{\mbox{{\tiny (parton)}}}
                                                L^{\mu \nu} \nonumber \\
       && ~~~~~~~~~~~\times
              \frac{d^3p_c}{(2\pi)^3 2E_c}
               \frac{d^3p_e}{(2\pi)^3 2E_e}
                \frac{d^3p_{\nu}}{(2\pi)^3 2E_{\nu}}~~,
\ee
where
\be
&&L^{\mu \nu} = 2~(p_e^{\mu} p_{\nu}^{\nu} + p_e^{\nu} p_{\nu}^{\mu}
                -g^{\mu \nu} p_e \cdot p_{\nu}
                +i \epsilon^{\mu \nu \alpha \beta}
                               {p_e}_{\alpha} {p_{\nu}}_{\beta})~~,
\nonumber \\
&&H_{\mu \nu}^{\mbox{{\tiny (parton)}}}
           = N~({p_b}_{\mu} {p_c}_{\nu} + {p_b}_{\nu} {p_c}_{\mu})~~.
\ee
Here $H_{\mu \nu}^{\mbox{{\tiny (parton)}}}$ denotes the partonic
subprocess'
hadronic tensor contrbuting $B \ra D$ decay.
In the inclusive decays $B \ra X_c e \nu$,
the hadronic tensor is given by
\be
W_{\mu \nu}=2~({p_b}_{\mu} {p_c}_{\nu} + {p_b}_{\nu} {p_c}_{\mu}
              -g_{\mu \nu} p_b \cdot p_c
          -i \epsilon_{\mu \nu \alpha \beta} p_b^{\alpha} p_c^{\beta})~~.
\nonumber
\ee
As can be easily seen, this hadronic tensor contains all the possible spin
configurations of $b \ra c$ transition from spin $0$ state.
Here we are interested in only $B \ra D$ process, where the spin does not
change
during the process. Therefore, we have to choose only the spin-inert part
which contributes to $B \ra D$ process from the inclusive hadronic tensor,
{}~$W_{\mu \nu}$.
By comparing with Eq. (14), we find that the spin-inert part
has the form of $({p_b}_{\mu} {p_c}_{\nu} + {p_b}_{\nu} {p_c}_{\mu})$,
as shown in (16).
Besides, we do not know how much spin-inert part
really contributes to $B \ra D$ semileptonic decay, because other decays
such as $B \ra D^*$, $B \ra D^{**}$ also contain spin-inert parts.
Therefore, the parameter $N$ is introduced to estimate the size of
spin-inert part which contributes to $B \ra D$ process.

Generally we can write the hadronic tensor of any exclusive semileptonic
decay modes for $B$ mesons as
\be
H_{\mu \nu}^{\mbox{{\tiny (parton)}}}
          = H_1(q^2)~({p_b}_{\mu} {p_c}_{\nu} + {p_b}_{\nu} {p_c}_{\mu})
           +H_2(q^2)~g^{\mu \nu} p_b \cdot p_c
           +i H_3(q^2)~\epsilon^{\mu \nu \alpha \beta}
                               {p_b}_{\alpha} {p_c}_{\beta}~~,
\ee
which is expressed in the form motivated by that of
inclusive $B \ra X_c e \nu$ decays.
With the expression of Eq. (17), we find
\be
H_1 = N,~~~~~~~~ H_2 = H_3 = 0. \nonumber
\ee
in the case of $B \ra D e \nu$ decay.
The constant $N$ will be later determined
by the zero recoil limit of the Isgur-Wise function.

Using the relation (5), we can write the hadronic tensor in the parton
level as follows
\be
H_{\mu \nu}^{\mbox{{\tiny (parton)}}}
           &=& N~xy~({p_B}_{\mu} {p_D}_{\nu} + {p_B}_{\nu} {p_D}_{\mu})~~.
\ee
The momentum conservation of the partonic  subprocess corresponds
to the momentum conservation in the hadronic level in our model,
as explained before.
So we can substitute the Dirac delta function $ \delta^4(p_b-p_c-q)$
for $ \delta^4(p_B-p_D-q)$ in Eq. (15) with no loss of generality.
Therefore, we write the decay width of $\bar{B} \ra De\bar{\nu}$,
\be
E_B \cdot d\Gamma (B \ra De\nu)
          &=& \int~dx~f_B(x)~f_D(y(x,q^2)) E_b~d\Gamma(b \ra ce\nu)
                \nonumber \\
          &=& \int dx~f_B(x) f_D(y(x,q^2))~
                (2\pi)^4 \delta^4(p_B-p_D-q)
                \nonumber \\
          &&~~~~~~~~\times {G_F}^2 |V_{cb}|^2
           ~N~xy(x,q^2)~({p_B}_{\mu} {p_D}_{\nu} + {p_B}_{\nu} {p_D}_{\mu})
                       L^{\mu \nu}
                \nonumber \\
          &&~~~~~~~~~~~~\times~y^2(x,q^2)~\frac{d^3p_D}{(2\pi)^3 2E_D}
                          \frac{d^3p_e}{(2\pi)^3 2E_e}
                          \frac{d^3p_{\nu}}{(2\pi)^3 2E_{\nu}}~~.
\ee
Hence we find the hadronic tensor
\be
H_{\mu \nu}(B \ra D e \nu)
         &=& N \int dx~f_B(x) f_D(y(x,q^2)) ~xy^3(x,q^2)~
               ({p_B}_{\mu} {p_D}_{\nu} + {p_B}_{\nu} {p_D}_{\mu})
               \nonumber \\
         &\equiv& N~{\cal F}(q^2)
               ({p_B}_{\mu} {p_D}_{\nu} + {p_B}_{\nu} {p_D}_{\mu})~~,
\ee
where we defined the function ${\cal F}(q^2)$ as
\be
{\cal F}(q^2) \equiv  \int dx~f_B(x) f_D(y(x,q^2)) ~xy^3(x,q^2)~~.
\ee
For given $q^2$ in our parton picture, the function ${\cal F}(q^2)$
measures the weighted transition amplitude, which is explicitly
given by the overlap integral
of distribution functions of initial and final state hadrons.

Comparing (20) with the Eq. (14), the Isgur-Wise function is calculated
within the parton model approach
\be
|\xi(v \cdot v')|^2 \left( 1 + {\cal O}(\frac{1}{m_Q}) \right)
                 = 4 N \cdot R \cdot {\cal F}(v \cdot v')~~.
\ee
As explained before, in our model all the non-perturbative QCD effects are
included through the distribution functions, so the function
${\cal F}(q^2)$ validly contains all $1/m_Q$ corrections.
{}From the value of the zero recoil limit of the Isgur-Wise function
with $1/m_Q$ corrections \cite{update}, we can determine the numerical
value of $N$. In Table 1, we show the numerical values
of $N$ with varying the parameters of the fragmentation functions,
($\epsilon_b$, $\epsilon_c$).
Unfortunately we cannot obtain the analytic form of
${\cal F}(q^2)$ due to the nontriviality of the integrals of $f(x)$.
However, we can numerically obtain the behaviors of the Isgur-Wise function
with varying $q^2$ from the Eq. (22),
and calculate its slope parameter.
The normalization constant $N$ and the slope parameters $\rho^2$
with the input values of ($\epsilon_b$, $\epsilon_c$)
are also shown in the Table 1.
We varied $\epsilon_b$ from 0.004 to 0.006, and
$\epsilon_c$ from 0.04 to 0.1.

The $q^2$ spectrum is given by
\be
\frac{d\Gamma(\bar{B} \ra D e \bar{\nu})}{dq^2}
  = \frac{{G_F}^2 |V_{cb}|^2}{96 \pi^3 m_B^3}
{\cal F}(q^2) \left( (m_B^2-m_D^2+q^2)^2-4 m_B^2 q^2 \right)^{3/2}~~.
\ee
And in Fig. 2(a), our prediction is shown
with the parameters ($\epsilon_b =$0.004,  $\epsilon_c =$0.04),
compared with Wirbel {\it et al}.'s
model prediction \cite{wsb}, which shows quite a good agreement.

\subsection{$B \ra D^* e \nu$}

We write the matrix element of the decay $\bar{B} \ra D^* e \bar{\nu}$
in familiar form
\be
<D^*|V_{\mu}+A_{\mu}|B>&&=
           \frac{2i}{m_B+m_{D^*}} \epsilon_{\mu \nu \alpha \beta}
           {\epsilon^*}^{\nu} p_{D^*}^{\alpha} p_B^{\beta} V(q^2)
        \nonumber \\
        + (m_B&&+m_{D^*}) {\epsilon^*}_{\mu} A_1(q^2)
    - \frac{\epsilon^* \cdot q}{m_B+m_{D^*}} (p_B+p_{D^*})_{\mu} A_2(q^2)
        \nonumber \\
        & & -2 m_{D^*} \frac{\epsilon^* \cdot q}{q^2} q_{\mu} A_3(q^2)
         +2 m_{D^*} \frac{\epsilon^* \cdot q}{q^2} q_{\mu} A_0(q^2)~~,
\ee
where
\be
A_3(q^2) = \frac{m_B+m_{D^*}}{2 m_{D^*}} A_1(q^2)
         - \frac{m_B-m_{D^*}}{2 m_{D^*}} A_2(q^2)~~, \nonumber
\ee
and in terms of the HQET
\be
<D^*(v')|V_{\mu}+A_{\mu}|B(v)> &&= \sqrt{m_Bm_{D^*}}
           \nonumber \\
         \times ( i \xi_V(&&v \cdot v') \epsilon_{\mu \nu \alpha \beta}
           {\epsilon^*}^{\nu} v'^{\alpha} v^{\beta} v(q^2)
           +\xi_{A_1}(v \cdot v')(v \cdot v'+1) {\epsilon^*}_{\mu}
           \nonumber \\
         &&-\xi_{A_2}(v \cdot v')\epsilon^* \cdot v v_{\mu}
           -\xi_{A_3}(v \cdot v')\epsilon^* \cdot v v'_{\mu})~~.
\ee
In fact, $A_0(q^2)$ and $A_3(q^2)$ do not
contribute here because of leptonic currents conservation.
With the heavy quark expansion, the hadronic form factors are related
to the Isgur-Wise function $\xi(v \cdot v')$, such as
$\xi_i(v \cdot v') = \xi(v \cdot v')\big( \alpha_i+{\cal O}(1/m_Q) \big)$,
where $\alpha_{A_2}=0$ and $\alpha_i=1$ otherwise.

The hadronic tensor of $\bar{B} \ra D^* e \bar{\nu}$ decay is obtained
using the HQET
\be
H_{\mu \nu} &=& \sum_{\mbox{{\tiny spin}}}
             <D^*| V_{\mu}+A_{\mu} |B> <D^*| V_{\nu}+A_{\nu} |B>^*
             \\
            &=& {R^*}^{-1} |\xi(v \cdot v')|^2
             \left[
	     \left(1-\frac{2q^2}{(m_B+m_{D^*})^2}\right)
	     ({p_B}_{\mu} {p_{D^*}}_{\nu} + {p_B}_{\nu} {p_{D^*}}_{\mu})
             (1+{\cal O}(\frac{1}{M_Q})) \right.
             \nonumber \\
      && \left.-2 \left( 1-\frac{q^2}{(m_B+m_{D^*})^2} \right)
             \left(g_{\mu \nu} p_B \cdot p_{D^*}
             (1+{\cal O}(\frac{1}{M_Q}))
          -i \epsilon_{\mu \nu \alpha \beta} p^{\alpha}_B p^{\beta}_{D^*}
             (1+{\cal O}(\frac{1}{M_Q})) \right)
             \right]~~,
             \nonumber
\ee
where $ R^*=2\sqrt{m_B m_{D^*}}/(m_B+m_{D^*})$. \nonumber
Investigating the above relation, we can find the form of the
hadronic tensor for the partonic subprocess which contributes to the decay
$\bar{B} \ra D^* e \bar{\nu}$.
The resulting hadronic tensor is written down in the form,
\be
H_{\mu \nu}^{(\mbox{{\tiny parton}})} &=&
                \left( 1-\frac{2q^2}{(m_B+m_{D^*})^2} \right) N_1~
	   ({p_b}_{\mu} {p_c}_{\nu} + {p_b}_{\nu} {p_c}_{\mu})
             \nonumber \\
          &&~~~~-2~\left( 1-\frac{q^2}{(m_B+m_{D^*})^2} \right)
           (N_2~g_{\mu \nu} p_b \cdot p_c
     -i N_3~\epsilon_{\mu \nu \alpha \beta} p^{\alpha}_b p^{\beta}_c)~~.
\ee
The parameters $N_i$'s give the relative size of form factors and
overall normalization. In general they are not constants and
have the $q^2$ dependences.
In the heavy quark limit, $N_i$'s become
constants and the values are equal to that of the normalization
constant $N$ in $B \ra D e \nu$ process.

In order to investigate the procedure more conveniently,
we define the ratios of form factors as follows:
\be
R_1 &\equiv& \left( 1-\frac{q^2}{(m_B+m_{D^*})^2} \right)
             \frac{V(q^2)}{A_1(q^2)}~~,
             \nonumber \\
R_2 &\equiv& \left( 1-\frac{q^2}{(m_B+m_{D^*})^2} \right)
             \frac{A_2(q^2)}{A_1(q^2)}~~,
\ee
where $V(q^2)$, $A_1(q^2)$ and $A_2(q^2)$ denote vector and axial vector
form factors respectively.
Then we can write the relations among form factors
and the Isgur-Wise function as
\be
A_1(q^2) &=& \left( 1-\frac{q^2}{(m_B+m_{D^*})^2} \right)
             {R^*}^{-1} \xi(q^2)~~,
             \nonumber \\
A_2(q^2) &=& R_2 {R^*}^{-1} \xi(q^2)~~,
             \nonumber \\
V(q^2) &=& R_1 {R^*}^{-1} \xi(q^2)~~.
\ee
Note that the Isgur-Wise function $\xi(v \cdot v')$
in these expressions contains full $1/m_Q$ corrections and
it corresponds to ${\hat \xi}(v \cdot v')$ in the Ref. \cite{update},
which is normalized at zero recoil up to the second order corrections
${\hat \xi}(1) = 1+\delta_{1/m_Q}$.
And $R_1$ and $R_2$ become to be unity in the heavy quark limit, and
Neubert's estimates of the $1/m_Q$ corrections
\cite{neubert} give
\be
R_1 \approx 1.3~~,~~~~~~~~R_2 \approx 0.8~~,
\ee
which are model dependent.
Recently CLEO \cite{cleo2} obtained the values of the parameters
$R_1$, $R_2$
by fitting them with the slope parameter of the Isgur-Wise function
$\rho^2$.
Since $R_1$ and $R_2$ have very weak $q^2$ dependence,
in the CLEO analysis they approximately obtained the  $q^2$
independent values,
\be
R_1 &=& 1.30 \pm 0.36 \pm 0.16~~, \nonumber \\
R_2 &=& 0.64 \pm 0.26 \pm 0.12~~.
\ee
Hereafter we also take them to be constants for simplicity.

In our model the parameters $N_1$, $N_2$ and $N_3$ are represented in
terms of $R_1$ and $R_2$ as follows,
\be
&&N_1 = \frac{N/2}{1-2q^2/(m_B+m_{D^*})^2}
        \left[
        -\frac{q^2}{(m_B+m_{D^*})^2} \cdot 2R_1^2 \right.
          \nonumber \\
&& \left.~~ +\left( 1-\frac{q^2}{(m_B+m_{D^*})^2}\right)
          \left(
          \frac{(1+R_2)^2}{2} + \frac{m_B^2+q^2}{2m_{D^*}^2} (1-R_2)^2
          +\frac{2m_B m_{D^*}-q^2}{2m^2_{D^*}} (1-R_2^2)
          \right)
        \right]~~,
        \nonumber \\
&&N_2 = \frac{N}{2} \left[
        (1+R_1^2) + \frac{2m_B m_{D^*}}{m_B^2+m_{D^*}^2-q^2} (1-R_1^2)
        \right]~~,
        \nonumber \\
&&N_3 = N R_1~~.
\ee
In the heavy quark limit, we know that $R_1=R_2=1$.
Using the expression $ R_i = 1 + {\cal O}(1/m_Q) $
we can separate the leading contributions and $1/m_Q$
corrections in $N_i$'s:
\be
N_1 &=& \frac{N}{4} (1+R_2)^2 + {\cal O}(1-R_2)~~,
        \nonumber \\
N_2 &=& \frac{N}{2} (1+R_1^2) + {\cal O}(1-R_1)~~,
        \nonumber \\
N_3 &=& N R_1~~.
\ee
When $R_1,~R_2 \ra 1$, we explicitly see that $N_1=N_2=N_3 \ra N$.

Now substituting $\xi(v \cdot v')$ for ${\cal F}(q^2)$ with the Eq. (22),
the hadronic tensor for the decay $\bar{B} \ra D^* e \bar{\nu}$
is given by
\be
H_{\mu \nu}(B \ra D^*) &=& {\cal F}(q^2)
         \left[
         \big(1-\frac{2q^2}{(m_B+m_{D^*})^2}\big) N_1
         ({p_B}_{\mu} {p_{D^*}}_{\nu} + {p_B}_{\nu} {p_{D^*}}_{\mu})
         \right.
             \nonumber \\
         &-& \left.2 \big(1-\frac{q^2}{(m_B+m_{D^*})^2}\big)
           (N_2 g_{\mu \nu} p_B \cdot p_{D^*}
           -i N_3 \epsilon_{\mu \nu \alpha \beta}
                     p^{\alpha}_B p^{\beta}_{D^*})
         \right]~~.
\ee
The $q^2$ dependences of form factors are mainly determined
by the function ${\cal F}(q^2)$, instead of the commonly used
pole-dominance ans\"atze.

When we calculate the hadronic tensor within the HQET framework,
we have generally some parameters parametrizing non-perturbative effects,
which are obtained in model dependent ways.
The slope parameter is such a characteristic parameter of
the Isgur-Wise function, which
represents the common behaviors of form factors.
We calculated it, and find that the value of the slope parameter
is related to the parameters $\epsilon_b$ and $\epsilon_c$
in our approach. The HQET also contains the parameter
$\lambda_a \sim -\langle k_Q^2 \rangle$
which is related to the kinetic energy of the heavy quark
inside the heavy meson, and spin-symmetry breaking term
$\lambda_2 = \frac{1}{4}(m_V^2-m_P^2)$
corresponding to the mass splitting of pseudoscalar mesons and vector mesons.
In our approach, we have two parameters $R_1$ and $R_2$ correspondingly,
which are experimentally measurable.
Also the values of $\epsilon_b$ and $\epsilon_c$ can be independently
determined from the various methods experimentally and theoretically.
We use the values for $R_1$ and $R_2$ from the CLEO
fit results of Ref. \cite{cleo2}, and for $\epsilon_b$ and $\epsilon_c$
from Ref. \cite{peterson,zphy}.

Finally, we obtain the decay spectrum
\be
\frac{d\Gamma(\bar{B} \ra D^* e \bar{\nu})}{dq^2}
  &=& \frac{G_F^2 |V_{cb}|^2}{192 \pi^3 m_B^5} {\cal F}(q^2)
      \left( (m_B^2-m_{D^*}^2+q^2)^2-4 m_B^2 q^2 \right)^{1/2}
      \nonumber \\
  & & ~~~
      \times \left[
        m_B^2 W_1(q^2)~((m_B^2-m_{D^*}^2+q^2)^2 -m_B^2 q^2) \right.
      \nonumber \\
  & & ~~~~~~~
      \left. + \frac{3}{2} m_B^2 W_2(q^2) ~(m_B^2-m_{D^*}^2+q^2)
                 + 3 m_B^2 W_3(q^2)
      \right]~~,
\ee
where
\be
W_1(q^2) &=& - N_1 \left( 1-\frac{2q^2}{(m_B+m_{D^*})^2} \right)~~,
        \nonumber \\
W_2(q^2) &=&  N_1 (m_B^2-m_{D^*}^2+q^2)
               \left(1-\frac{2q^2}{(m_B+m_{D^*})^2}\right)
       - 2 N_3 q^2 \left( 1-\frac{q^2}{(m_B+m_{D^*})^2} \right)~~,
        \nonumber \\
W_3(q^2) &=& - N_1 m_B^2 q^2 \left( 1-\frac{2q^2}{(m_B+m_{D^*})^2} \right)
        \nonumber \\
    & &~~~~+ N_3 q^2 (m_B^2-m_{D^*}^2+q^2)
         \left( 1-\frac{q^2}{(m_B+m_{D^*})^2} \right)
        \nonumber \\
    & &~~~~~~~~+ N_2 q^2 (m_B^2+m_{D^*}^2-q^2)
         \left( 1-\frac{2q^2}{(m_B+m_{D^*})^2} \right)~~,
\ee
and ${\cal F}(q^2)$ is defined in (21).
The result is plotted in Fig. 2 (b), also
compared  with the CLEO data \cite{cleo2}.
The thick solid line is our model prediction
with the parameters ($\epsilon_b =$0.004,  $\epsilon_c =$0.04),
the thin solid line the Wirbel {\it et. al.} model prediction \cite{wsb},
and the dotted line the K\"orner {\it et. al.} model prediction \cite{ks}.

\section{Discussions and Conclusions}

All form factors show the same behavior for
varying $q^2$, which is described by the Isgur-Wise
function of the HQET, which represents the common non-perturbative
dynamics of weak decays of heavy mesons.
Ever since Fakirov and Stech \cite{fakirov}, the
nearest pole dominance has been usually adopted as the dependence of
common behaviors on $q^2$.
In our approach, their $q^2$ dependences are  {\bf derived}
from the kinematic relations of $b$- and $c$-quark.
When $b$-quark decays to $c$-quark, the momentum transfer
to leptonic sector is equal to
the difference between $b$-quark momentum and $c$-quark momentum
in the parton picture.
The $b$- and $c$-quark momenta within the $B$ and $D$ mesons
have some specific distributions.
For given momentum transfer $q^2$, there exist possible
configurations of $b$- and $c$-quark momentum pairs $(p_b,~p_c)$,
and each pair is appropriately weighted with the momentum distributions
of the quarks.
Our ${\cal F}(q^2)$ function in (21) measures the weighted transition
amplitude for given $q^2$ in the parton picture; it is explicitly
given by the overlap integral
of distribution functions of initial and final state hadrons.
This is common to all form factors, as explained in Section II.

As mentioned earlier, all non-perturbative strong interaction
effects are considered through the distribution functions
in our model, so $\xi(v \cdot v')$ obtained from Eq. (22)
corresponds to the hadronic form factor ${\hat \xi}(v \cdot v')$
defined in the Ref. \cite{update}, including $1/m_b$ corrections
rather than the lowest order Isgur-Wise function.
And the slope parameter of our results in Table 1 also corresponds to
${\hat \rho}^2$ related to ${\hat \xi}(v \cdot v')$.
We obtain the values of the slope parameter $\rho$ within the
parton model framework, as in Table 1,
\be
 \rho^2 = 0.582 - 0.896~~,
 \nonumber
\ee
which are compatible with the Neubert's prediction \cite{update},
\be
{\hat \rho}^2 \simeq \rho^2 \pm 0.2 = 0.7 \pm 0.2~~.
 \nonumber
\ee
Our result is  rather smaller than the predictions of other models,
\be
 \rho^2 &=& 1.29 \pm 0.28 ~~~~~~\mbox{\cite{rosner}}~~,
 \nonumber \\
 \rho^2 &=& 0.99 \pm 0.04 ~~~~~~\mbox{\cite{mannel}}~~,
 \nonumber
\ee
but it is consistent with the average value measured
by experiments \cite{exps},
\be
 {\hat \rho}^2 = 0.87 \pm 0.12~~.
 \nonumber
\ee

In calculating the numerical values, we still have two free
parameters $\epsilon_b$ and $\epsilon_c$ of Eq. (6),
{\it i.e.} of the heavy quark fragmentation functions.
Their values can be determined
independently from the various experimental and theoretical
methods\footnote{Theoretically, it is interesting to calculate
$f_Q(z)=\delta(1-z)+{\cal O}(1/m_Q^2)$ within the HQET, which can reproduce
the phenomenological predictions.}.
For the parton model to be consistent with the HQET,
we require that with the fixed value of the parameter $\epsilon_Q$,
all the appropriate results of the parton model approach agree
with those of the HQET.
In other words, the value of parameter $\epsilon_Q$ should be determined
to give all the phenomenological results to coincide
with those of the HQET.
In this point of view, we have previously studied the parton model
approach
for inclusive semileptonic decays of $B$ meson in the Ref. \cite{ky},
and showed that the value $\epsilon_b \approx 0.004$
gives consistent results with those of the HQET.
In this paper we use the value of $\epsilon_b$ as 0.004
or 0.006. The latter value is given by the experiments for the
determination of the Peterson fragmentation function \cite{zphy}.
We find that our prediction of the slope parameter $\rho^2$
with the parameter $\epsilon_b = 0.004$ and $\epsilon_c = 0.04$
gives the best agreed value 0.705 with that of the HQET,
${\hat \rho}^2 = 0.7 \pm 0.2$. In this context,
we conclude that our model with the parameter $\epsilon_b = 0.004$
gives consistent predictions with the HQET.

Phenomenologically, our model prediction on $q^2$
spectrum in the $B \ra D^* e \nu$ decay shows a good
agreement with the result of the CLEO \cite{cleo2}, as shown in Fig. 2(b).
If we let $f_Q(z)=\delta(1-z)$ and $R_1=R_2 =1$ in our model,
we can reproduce the lowest order results of the HQET, and
obtain the similar plot with those of other models in Fig. 2(b).
For the $B \ra D e \nu$ decay, our results agrees with those
of other models, as in Fig. 2(a).

Finally we obtain the ratio of integrated total widths
$\Gamma (B \ra D^*)/ \Gamma (B \ra D) \approx 2.66$,
which agrees with the experimental results \cite{pdg}.
It may be a phenomenological support of our model
because this quantity is independent of the CKM elements $|V_{cb}|$,
which has uncertainties in determining its value yet.
The perturbative QCD corrections can be factorized in the decay width
calculation \cite{update,pert},
which does not affect the ratio $\Gamma (B \ra D^*)/ \Gamma (B \ra D)$.

To summarize, we developed the new parton model approach for
exclusive semileptonic decays of $B$-meson by
extending the inclusive parton model,
and by combining with the results of the HQET, motivated by
Drell-Yan process.
Without the nearest pole dominance ans\"atze, we derived the
dependences of hadronic form factors on $q^2$. We also calculated
numerically the slope of the Isgur-Wise function, which is consistent
with the experimental results.

\acknowledgements

We thank Pyungwon Ko and E. Paschos for their careful reading
of manuscript and their valuable comments.
The work was supported
in part by the Korean Science and Engineering  Foundation,
Project No. 951-0207-008-2,
in part by Non-Directed-Research-Fund, Korea Research Foundation 1993,
in part by the CTP, Seoul National University,
in part by Yonsei University Faculty Research Grant 1995,  and
in part by the Basic Science Research Institute Program, Ministry of
Education, 1994,  Project No. BSRI-94-2425.

%
\begin{table}
\begin{center}
\vspace{2cm}
\caption{
The normalization constant $N$ and the slope parameter $\rho$
are shown with several values of the parameters
$\epsilon_b$,
$\epsilon_c$.
}
\vspace{4cm}
\begin{tabular}{cccccccccccccr}
&&&&&&&&&&&&&\\
& & & & $\epsilon_b =$ & 0.004 & & & & $\epsilon_b =$ & 0.006 & & &  \\
& & & $\epsilon_c =$ 0.04 & 0.06 & 0.08 &0.1 & &
0.04 & 0.06 & 0.08 & 0.1 &\\
\hline
&&&&&&&&&&&&&\\
& $N$ & & 0.938 & 1.068 & 1.190 & 1.306 & & 1.105 & 1.223 & 1.338 & 1.449 &
\\
&&&&&&&&&&&&&\\
& $\rho^2$ & & 0.705 & 0.646 & 0.609 & 0.582 & & 0.896 &
0.844 & 0.810 & 0.785 &
\\
&&&&&&&&&&&&&\\
\end{tabular}
\end{center}
\end{table}
%
%

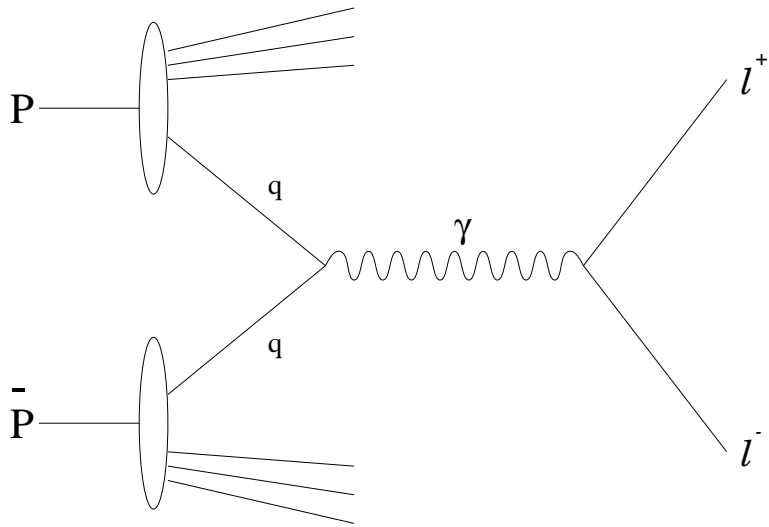
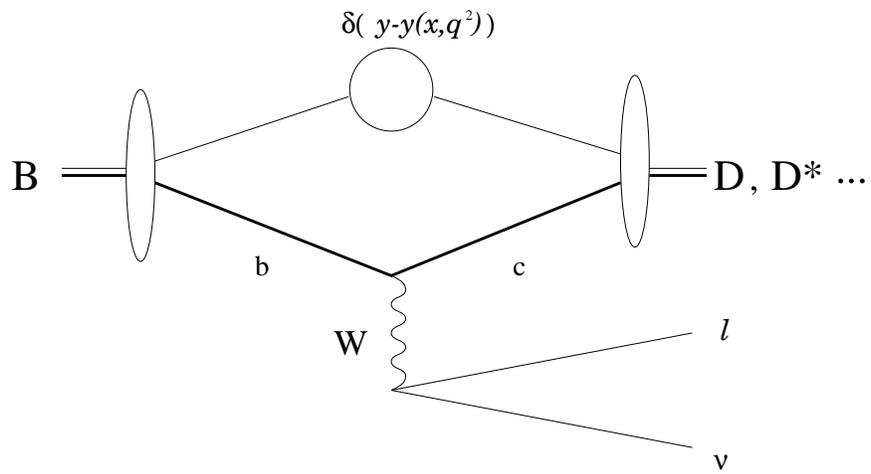
\begin{figure}
\caption{
(a) The diagram of Drell-Yan Process. (b) The schematic diagram of
semileptonic exclusive decay of $B$-meson in the parton model.  }
\end{figure}

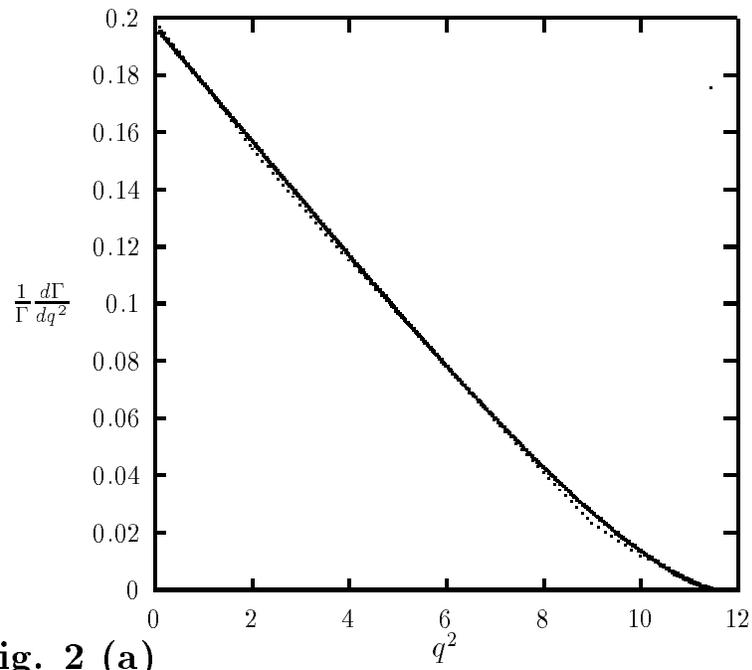
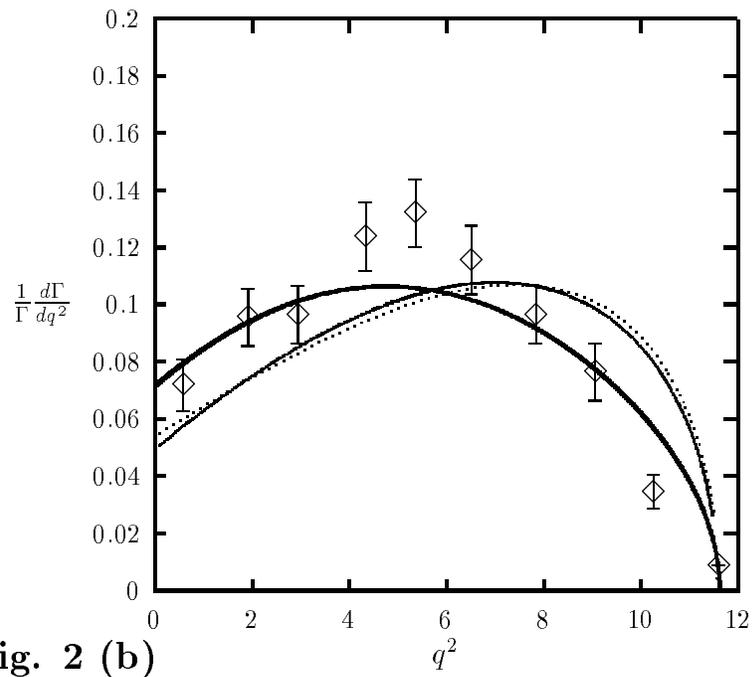
\begin{figure}
\caption{
(a) $q^2$ spectrum in the $B \ra D e \nu$ decays.
The solid line is our model prediction and
the dotted line the Wirbel {\it et. al.} model prediction \protect\cite{wsb}.
We used the values of parameters,
($\epsilon_b =$0.004,  $\epsilon_c =$0.04).
(b) $q^2$ spectrum in the $B \ra D^* e \nu$ decays.
The thick solid line is our model prediction and
the thin solid line the Wirbel {\it et. al.}
model prediction \protect\cite{wsb} and
the dotted line the K\"orner {\it el. al.}
model prediction \protect\cite{ks}.
We used the values of parameters,
($\epsilon_b =$0.004,  $\epsilon_c =$0.04).
The data are quoted from Ref. \protect\cite{cleo2}.
}
\end{figure}

\end{document}